\documentclass[prl,twocolumn,amsmath,aps,floats,amssymb, floatfix, superscriptaddress, footinbib, preprintnumbers,10pt]{revtex4-1}

\usepackage{graphicx}
\usepackage{wasysym}
\usepackage{amsfonts}
\usepackage{subfigure}

\def\beq{\begin{equation}}
\def\eeq{\end{equation}}
\def\be{\begin{equation}}
\def\ee{\end{equation}}
\def\bea{\begin{efqnarray}}
\def\eea{\end{eqnarray}}

\def\mpl{M_{\rm Pl}}

\begin{document}

\preprint{IPMU-10-0163}

\title{Identifying the inflaton with primordial gravitational waves}
\medskip\
\author{ Damien A. Easson}%
\email[Email:]{easson@asu.edu}
\affiliation{ Department of Physics  \& School of Earth and Space Exploration  \& Beyond Center,
Arizona State University, Tempe, AZ 85287-1504, USA}
\affiliation{Institute for the Physics and Mathematics 
of the Universe, University of Tokyo,
5-1-5 Kashiwanoha, Kashiwa, Chiba 277-8568, Japan}
\author{Brian A.\ Powell} 
\email[Email:]{brian.powell007@gmail.com}
\affiliation{Institute for Defense Analyses, Alexandria, VA 22311, USA}

\begin{abstract}
We explore the ability of experimental physics to uncover the underlying structure of the gravitational Lagrangian describing inflation. While the observable degeneracy of the inflationary
parameter space is  large, future measurements of observables beyond the adiabatic and tensor two-point functions, such as non-Gaussianity or isocurvature modes, might reduce this degeneracy. We show that even in the absence
of such observables, the range of possible inflaton potentials can be reduced with a precision measurement of the tensor spectral index, as might be possible with a direct detection of
primordial gravitational waves.
\end{abstract}

\maketitle
\noindent

\large{I}\normalsize n the simplest realizations of the inflationary universe paradigm, acceleration expansion is generated by a single canonically normalized scalar inflaton field $\phi$ with potential $V(\phi)$. Within this setting, there
exists a unique mapping between the set of observables and the free parameters of the Lagrangian; in single field, slow roll inflation the observables are
determined by the potential, $\mathcal{O}[V(\phi)]$.  
There are, however, a vast number of ways to achieve the desired acceleration including
models involving multi-fields, non-standard kinetic terms, and non-trivial gravitational couplings.
In such elaborate settings the observables depend on modified or even additional degrees of freedom, $\mathcal{O}[V(\phi),F(\phi,\cdots)]$. Further, any process designed to `invert' a subset of observables to
obtain the underlying free parameters of the Lagrangian will reveal a space of
Lagrangians that is not observationally unique. 
This \it degeneracy problem \rm is well-known and is a formidable challenge for cosmologist attempting to identify the inflaton.
Fortunately, many of the alternatives
to canonical single field inflation produce unique observational signatures, such as non-Gaussian and/or isocurvature
perturbations.  Such observations beyond the two-point adiabatic and tensor power spectra can be used to
distinguish between these models and break the degeneracy \cite{note0}.  However, of particular concern is the future
possibility that such discriminating observables are not detected.   While this problem has been previously
documented, there has been no attempt to systematically           
determine the size of the degeneracy, for example, by estimating the envelope of different potentials, $V(\phi)$,         
within the larger class of inflation theories that yield the same observables.
\begin{figure}
\centerline{\includegraphics[width=2.5in]{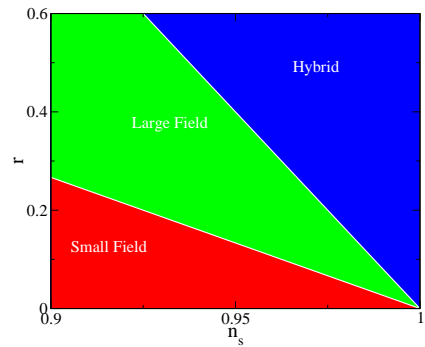}}
\caption{\scriptsize{Zoology of inflation models.}}
\end{figure}

In this work, we take an initial step in estimating the magnitude of this degeneracy by
performing Monte Carlo potential reconstructions within the context of two broad classes of alternatives to canonical single
field inflation, in the absence of discriminating observations. First we consider the case in which
the perturbation spectra are generated
by degrees of freedom that are decoupled from the inflationary dynamics, and second, 
the case where the inflationary dynamics are extended beyond
the paradigm of single field, canonical inflation by altering dynamical degrees of freedom.
As representative examples of such models we examine, respectively: the curvaton scenario, in which a non-inflationary degree of
freedom generates the primordial perturbations,  and DBI inflation, in which a non-canonical
kinetic term contributes to the inflationary dynamics.

%

In canonical single field models, the reconstruction program reveals that inflationary
potentials can be grouped into  three distinct
classes based on their observable predictions for $n_s$, the spectral index of
scalar perturbations, and $r$, the tensor/scalar ratio, i.e., 
vast numbers of inflationary
potentials organize themselves into a few observational families  (c.f. Figure 1). This classification scheme is commonly referred to as the `zoology' of inflation models \cite{Dodelson:1997hr,Kinney:2003uw}.  
`Hybrid' models include potentials that evolve asymptotically to their minima,
requiring an auxiliary field to end inflation.  However, they are effectively single field models with
non-vanishing energy density at the minimum, and have the common form $V(\phi) \propto 1 + (\phi/\mu)^p$,
where $\mu$ is an energy scale and $p$ a positive integer.  They are characterized by the conditions $V''(\phi) > 0$ and $({\rm log}V(\phi))'' > 0$.  The simplest models of tree-level hybrid
inflation \cite{Linde:1993cn,Copeland:1994vg} belong to this class.  `Small field' and `large field' models are differentiated by
their initial field values.  
Large field models, for example $m^2\phi^2$ inflation \cite{Linde:1983gd}, are characterized by a field initially
displaced far from its minimum, with the general form $V(\phi) \propto (\phi/\mu)^p$, satisfying $V''(\phi) > 0$ and $({\rm log}V(\phi))'' < 0$.  Conversely, small
field models are characterized by a field initially close to the origin, with general form $V(\phi) \propto 1 - (\phi/\mu)^p$ (near the maximum), satisfying $V''(\phi) < 0$ and $({\rm log}V(\phi))'' < 0$; `new' inflation and
models based on spontaneous symmetry breaking belong to this class.  
If the simplest implementation
of canonical single field inflation is assumed, we may hope to determine which class of the above potentials 
is ultimately responsible for driving inflation. 
To what degree is our ability to reconstruct the
physics of inflation threatened by relaxing this assumption?  
\begin{figure*}
\label{MC}
$\begin{array}{cc}
\subfigure[]{
\includegraphics[width=0.3 \textwidth,clip]{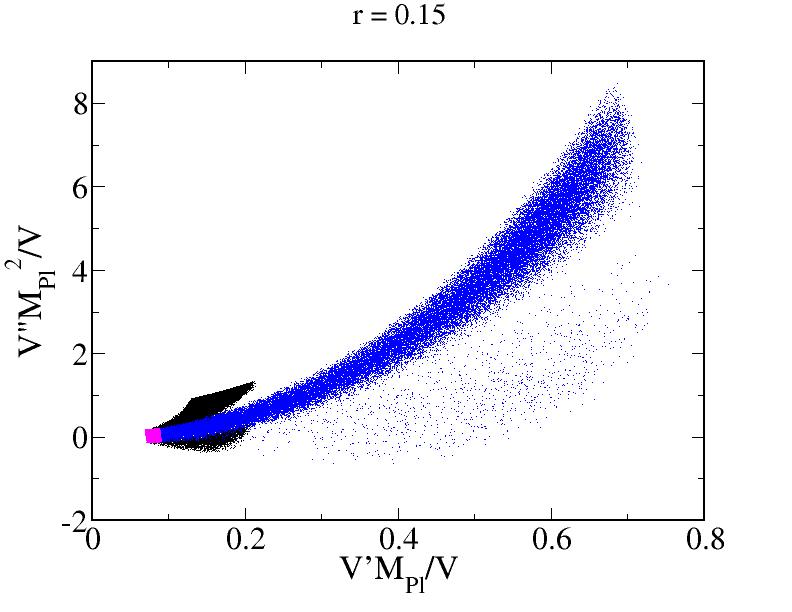}}
\hspace{0.5in}
\subfigure[]{
\includegraphics[width=0.29 \textwidth,clip]{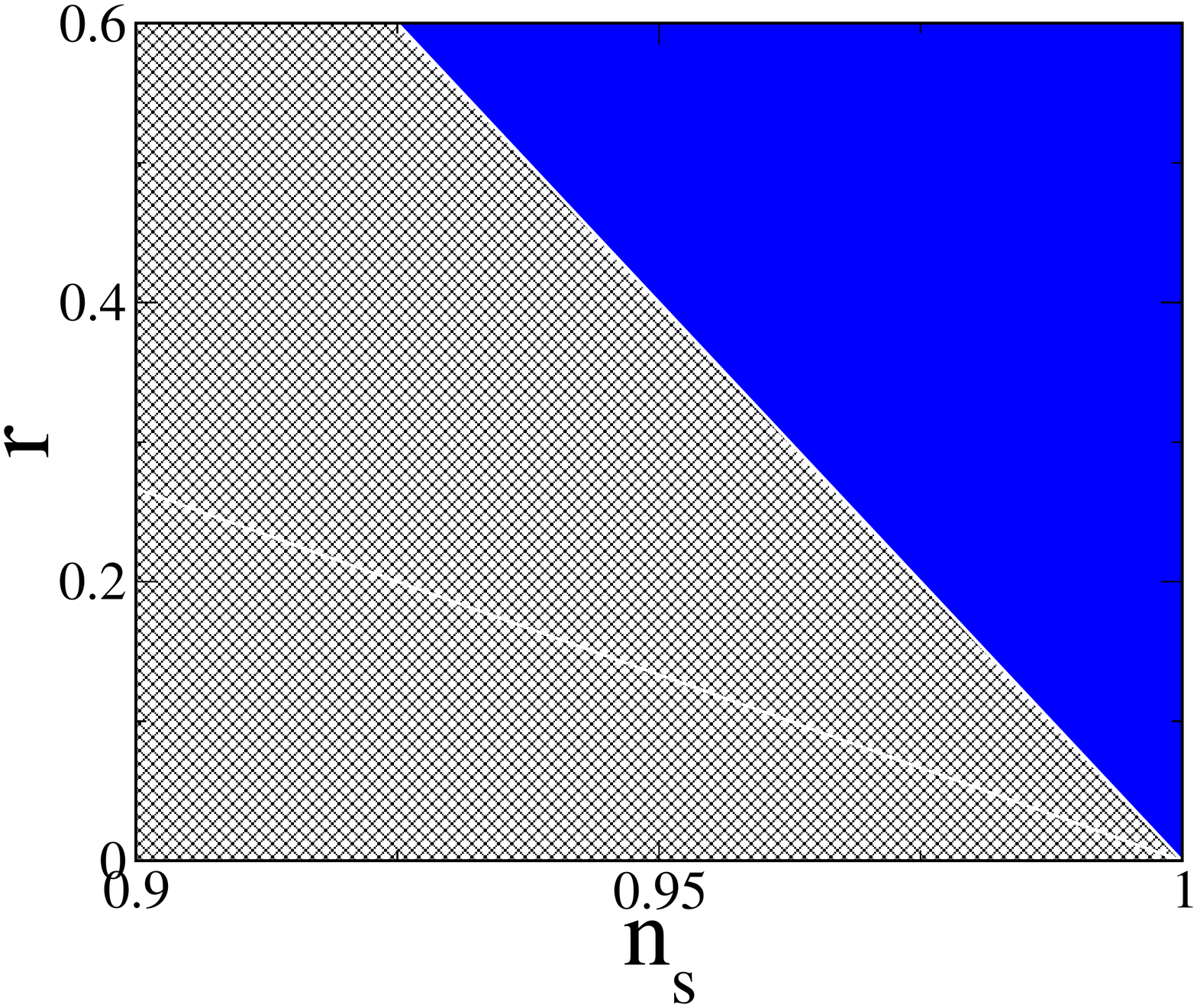}}
\end{array}$\\
$\begin{array}{cc}
\subfigure[]{
\includegraphics[width=0.3 \textwidth,clip]{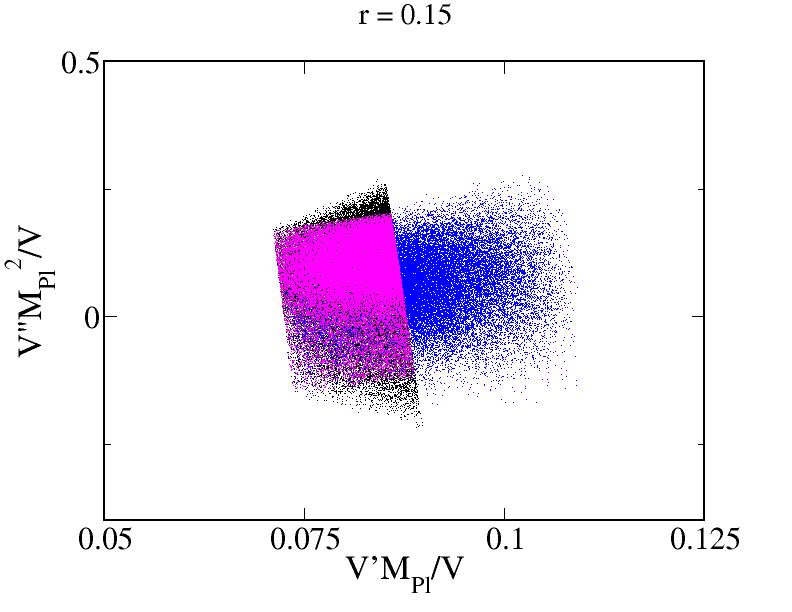}}
\hspace{0.5in}
\subfigure[]{
\includegraphics[width=0.3 \textwidth,clip]{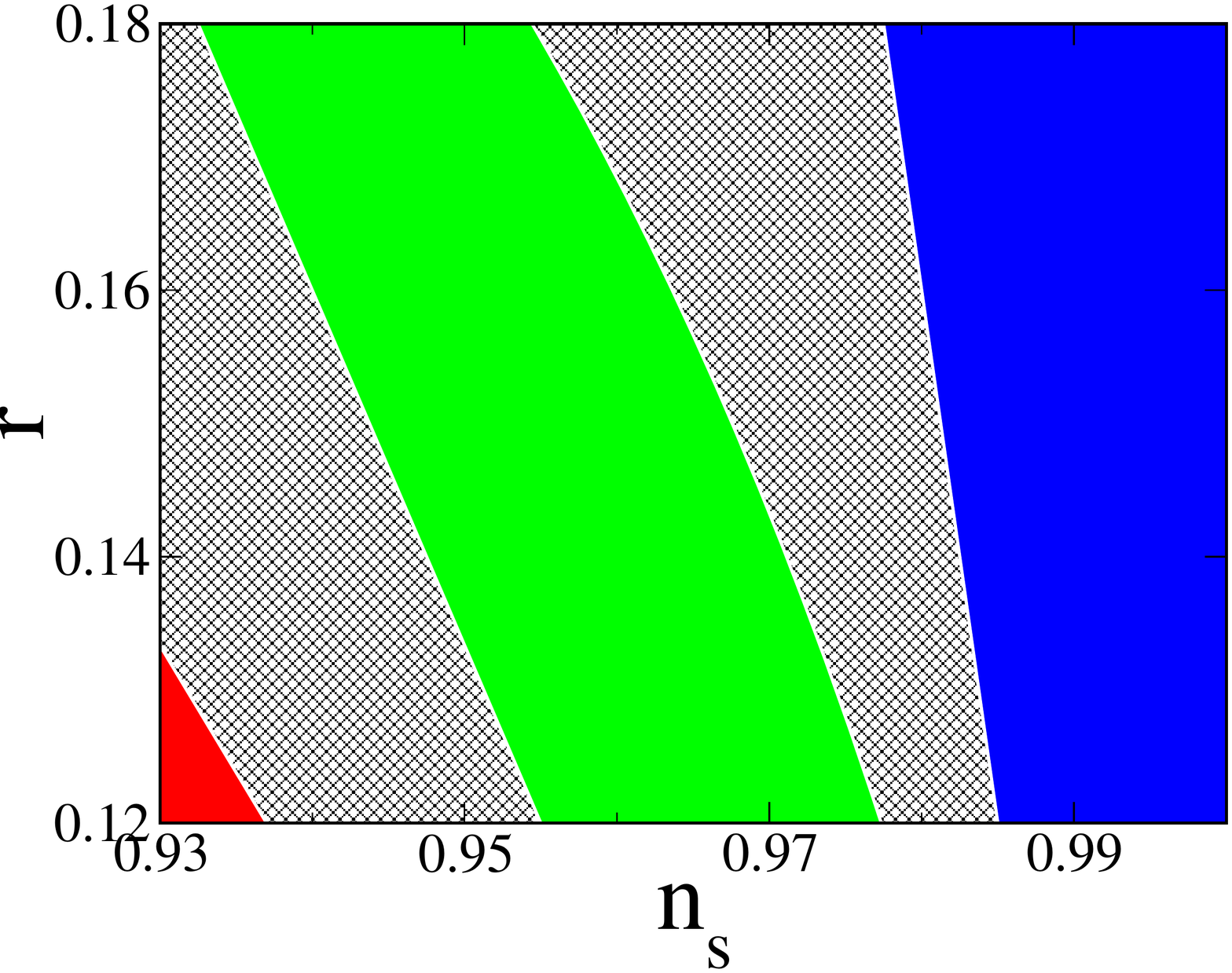}}
\end{array}$
\caption{\footnotesize{(a): Monte Carlo results for the worst-case degeneracy at Planck-precision using $r$, $n_s$, and
$dn_s/d{\rm ln}k$ in the canonical single field (magenta), curvaton (black) and DBI
(blue) reconstructions. (b): Zoology of the worst-case degeneracy (a).  Gray areas denote regions in which multiple classes overlap:
only hybrid models can be uniquely classified.
(c): Monte Carlo results for the best-case degeneracy utilizing a direct detection of primordial gravitational waves at DECIGO
precision to constrain $n_T$ with the same observables as in (a). 
(d): Zoology of models that results from the best-case degeneracy (c).}}
\end{figure*}
%
%

In the curvaton scenario  \cite{Linde:1996gt,Lyth:2001nq,Moroi:2001ct}, 
the central assumption of traditional reconstruction -- that the inflaton generates the
primordial spectra -- is relaxed.  Without knowledge of how the spectra were generated,
whether by the inflaton or by some other means, a unique inversion of observables is
clearly impossible.
The curvaton field, $\sigma$, is weakly coupled and relatively light during
inflation, $m^2 \ll H^2$. It 
influences the primordial power spectrum, $P_\Phi(k) = k^3|\Phi|^2/2\pi^2$, via the final curvature perturbation  
\begin{equation}
\label{pert1}
\Phi = -\frac{1}{2}\frac{H}{M_{\rm Pl}^2H'}\delta \phi - \frac{\tilde{f}(\sigma)}{\sqrt{2}M_{\rm Pl}}\delta \sigma,
\end{equation}
where $\delta \phi $ and $\delta
\sigma$ are the inflaton and curvaton vacuum fluctuations, and $\tilde{f}(\sigma)$ controls the
contribution of the curvaton to the overall perturbation. After inflation ends, the curvaton rolls to its
minimum where it begins to oscillate during the post-inflationary phase. These oscillations set up a small isocurvature perturbation that grows with time.
After the curvaton decays, the perturbation is converted to an adiabatic mode and structure begins to evolve according to the standard model. Measurements of
the adiabatic perturbation spectrum derived from Eq. (\ref{pert1}) and the tensor/scalar ratio $r$  do not uniquely determine the potential $V(\phi)$ unless
$\tilde{f}(\sigma)$ can be constrained, leading to the possibility of the same
potential giving rise to a wide range of observables.  For example, the first two
derivatives of $V(\phi)$ can be written,
\begin{eqnarray}
\label{vcurv}
V_{\rm curv}' &=& \sqrt{2}\frac{V_0}{M_{\rm Pl}}\left(\frac{r}{16-\tilde{f}^2r}\right)^{1/2},\\ 
\label{vcurv2}
V_{\rm curv}'' &=& \frac{V_0}{M_{\rm Pl}^2}\left[\frac{8(n_s-1) +
3r}{16-\tilde{f}^2r}\right].
\end{eqnarray}
However, depending on the thermal history and the energy density of the curvaton at the time
of decay, there may be residual isocurvature modes or primordial `local'-type non-Gaussianity large enough to be
detected in future experiments; these additional observables might enable a determination of $\tilde{f}(\sigma)$.
We consider the effects of such observations on reconstruction in \cite{debp10a} -- in this analysis we assume that
they are \it not \rm detected.

A degeneracy problem might also arise in the context of non-canonical models
\cite{debp10b,Powell:2008bi}.
In non-canonical models, the inflaton field Lagrangian includes non-standard
kinetic terms $\mathcal{L}(X, \phi)$, where, $2 X \equiv \partial^\mu \phi \partial_\mu \phi$. We study the
most well-motivated case -- that of the non-linear Lorentz invariant DBI action 
$\mathcal{L} = -f^{-1}(\phi)\sqrt{1+2f(\phi)X}+f^{-1}(\phi)-V(\phi)$, where $f(\phi)$ is the ``warp factor".
In the DBI model, the inflaton speed is bounded from above by a generalized Lorentz factor
$\gamma^{-1} \equiv \sqrt{1-f(\phi)\dot{\phi}^2}$, which can lead to a new type of
slow roll inflation even with
steep potentials. As a result, cosmological fluctuations travel with sound speed less than unity, $c_s = \gamma^{-1} \leq 1$,
leading to a curvature perturbation that depends on $\gamma$,
\begin{equation}
\Phi = -\frac{1}{2}\frac{H\gamma}{\mpl^2 H'}\delta \phi.
\end{equation}
Despite the formal distinction between the curvaton and DBI reconstructions, the treatment of the two
cases is the same: a determination of $V(\phi)$ requires observations of more than simply the adiabatic
density perturbation and tensor/scalar ratio.  The potential in DBI inflation gives  
\begin{eqnarray}
\label{vdbi}
V_{\rm DBI}' &=& \frac{V_0}{M_{\rm Pl}}\sqrt{\frac{r}{8}}\gamma,\\
\label{vdbi2}
V_{\rm DBI}'' &=& \frac{V_0}{2M_{\rm Pl}^2}\gamma\left(n_s-1 + \frac{3}{8}r\gamma\right).
\end{eqnarray} 
In the case of curvatons, the function $\tilde{f}(\sigma)$
needs to be measured; in the case of DBI inflation, the $\gamma$-factor must be constrained.  While large equilateral non-Gaussianities might be produced in DBI inflation, we
assume that future missions fail to detect them.

We consider only the minimal set of observational parameters
describing the primordial scalar and tensor power spectra: $P_\Phi(k)$ and $r$.  
To ascertain the size of the degeneracy, we employ the flow formalism \cite{Hoffman:2000ue,Kinney:2002qn}, which is a Monte Carlo approach to potential
reconstruction \cite{Easther:2002rw}.  The inflationary model space is stochastically
sampled and models of interest can be selected out.  We first seek to determine the constraints that can be imposed on $V(\phi)$ at Planck-precision \cite{note2},  in
the absence of discriminating observations: we consider 68\% CL detections of  $r$ ($r \gtrsim 0.01$, $\Delta r \sim 0.03$) \cite{Colombo:2008ta}, $n_s$ ($\Delta
n_s \sim 0.0038$), and $dn_s/d{\rm ln}k$ ($\Delta dn_s/d{\rm ln}k \sim 0.005$)
\cite{Bond:2004rt}.  Since $\Delta n_T \sim 0.1$ with a Planck B mode detection, the tensor spectral index will not
be well-resolved and will not be included in the reconstruction.  This worst-case reconstruction therefore makes
use of only the adiabatic and tensor two-point functions on CMB scales.
We perform
separate analyses for curvatons, DBI, and canonical single field inflation.  We collect only models that support at least 10 e-foldings of inflation
and satisfy the above observational constraints at  $k = 0.01 {\rm Mpc}^{-1}$.  We present the constraints on the
first two derivatives of $V(\phi)$ in Figure 2a: magenta, black, and blue points denote single field, curvaton, and
DBI models, respectively. The constraints depend only weakly on the fiducial observables chosen \cite{debp10a}: in Figures 2a and
2c we
choose $r = 0.15$, $n_s = 0.97$, and $dn_s/d{\rm ln}k = 0$ for the potential reconstructions.  If $r$ is not measured ($r\lesssim 0.05$ with Planck) then the uncertainty in
$V(\phi)$ extends to $V'/V = 0$, but is of the same order of magnitude as in the case $r=0.15$ \cite{debp10a,debp10b}.
Even an improved measurement of $r$ by next-generation CMB experiments like CMBPol will scarcely improve constraints on $V(\phi)$ in the presence of the degeneracy
\cite{debp10a,debp10b}. 

We next examine the effects of the unresolved degeneracy on the zoology by sorting the curvaton and DBI models by $(n_s,r)$ according to
the potential classification: small field, large field, and hybrid. We find that all observables that are compatible with
canonical single large field models are
also consistent with curvaton and/or DBI hybrid models. Furthermore, we find that all observations compatible with canonical
single small field models are also consistent with both large field and hybrid
curvaton and/or DBI models.
Only those hybrid models existing in the single field `hybrid' region can be
correctly classified in the presence of the degeneracy, i.e. they must  satisfy $r > 8(1 -n_s)$.
We present the zoology in Figure 2b indicating in gray regions in which at least two classes of model overlap.  

We have obtained the worst-case degeneracy by utilizing only the two-point adiabatic and tensor spectra on CMB scales
in the reconstructions.  It is certainly possible that these will be the only detected observables: canonical single field 
inflation could be the true underlying model, curvatons need not generate detectable
isocurvature modes or non-Gaussianity, and DBI inflation will fail to generate observable non-Gaussianity if the sound speed
$c_s \gtrsim 0.1$.  However, we need not restrict ourselves to observables on CMB scales only: primordial gravitational waves on
scales $k_* \sim 10^{14}$ Mpc$^{-1}$ can be used to measure the tensor spectral index, $n_T$, at a precision surpassing that
possible with a detection of B modes on CMB scales.  Future space-based laser interferometers, like Big Bang Observer (BBO) \cite{bbo} and Japan's Deci-hertz Interferometer Gravitational Wave Observatory           
(DECIGO) \cite{Seto:2001qf}, will detect gravitational waves if B modes on CMB scales give $r \gtrsim 10^{-3}$ and $r \gtrsim           
10^{-6}$, respectively.  This range includes a substantial portion of the inflationary model space. 
In comparison with an ideal                                                                                           
B mode detection on CMB scales, a direct detection with BBO will yield comparable constraints ($\Delta n_T \sim 10^{-2}$)
while 
DECIGO gives the best measurement: $\Delta n_T \sim 10^{-3}$ or better \cite{Seto:2005qy,Kudoh:2005as}. 

The tensor index turns out to be highly valuable to the reconstruction program, since, while canonical single field inflation predicts the
consistency condition $r = -n_T/8$, alternative theories typically yield modified relations: curvatons predict $r=-16n_T/(2-\tilde{f}^2(\sigma)n_T)$ and DBI inflation predicts         
$r = -8c_sn_T$.  With the modified conditions, we find that $\tilde{f}^2(\sigma)$ drops out of the curvaton reconstruction, Eqs.
(\ref{vcurv}-\ref{vcurv2}), giving $V_{\rm curv}' \propto V_0 \sqrt{-n_T}$ and $V_{\rm curv}'' \propto -V''_{\rm csf}n_T/r$, where $V''_{\rm
csf}$ is the canonical
single field reconstruction.  Likewise for DBI, $\gamma$ vanishes from Eqs. (\ref{vdbi}-\ref{vdbi2}) giving $V_{\rm DBI}' \propto -V_0 n_T/\sqrt{r}$ and $V_{\rm
DBI}'' \propto -V_0(n_s
-1 -3n_T)n_T/r$.  

We stress that we are not considering cases in which the values of $r$ and $n_T$ violate one or more of the above consistency
conditions, \it i.e. \rm we are assuming that the degeneracy remains intact.
The range of $V(\phi)$ is reduced despite the unbroken degeneracy because the consistency conditions constrain precisely the degrees of freedom
that are necessary for an inversion of the potential: $\tilde{f}^2(\sigma)$ for curvatons and $c_s$ for DBI inflation.  We need
not know \it a priori \rm which condition to impose -- we impose each one that agrees with the fiducial $r$ and $n_T$ to within
experimental error and perform the reconstruction.

We assume that the tensor spectrum is of the form
\begin{equation}                                                                                                                        
\label{tp}
P_h(k) = P_h(k_0)\left(\frac{k}{k_0}\right)^{n_T + \frac{1}{2}\alpha_T{\rm ln}\left(\frac{k}{k_0}\right)},                              
\end{equation}
where $\alpha_T = dn_T/d{\rm ln}k$ is the tensor index running and $k_0 = 0.01$ Mpc$^{-1}$.  The challenge is that direct
detection experiments determine $n_T(k_*)$, while the consistency relations are functions of $n_T(k_0)$.  In principle, the spectrum Eq.
(\ref{tp}) provides the mapping from $k_*$ to $k_0$; however, $\alpha_T$ is unlikely to be reliably constrained
by these experiments.  Although likely small, our ignorance of $\alpha_T$ limits the accuracy of the extrapolation from $k_0$ to $k_*$ and contributes to the error on $n_T$
\cite{Smith:2005mm,Smith:2006xf}, 
\begin{equation}                                                                                                                        
\label{errornt}                                                                                                                         
\Delta n_T = \left\{\left[\frac{6\times 10^{-18}}{XA_{GW}P_h(k_*)}\right]^2 + \left[\frac{1}{2}\alpha_T{\rm
ln}\left(\frac{k_*}{k_0}\right)\right]^2\right\}^{1/2},                                                                                 \end{equation}
where $A_{GW}  = 2.74 \times 10^{-6}$ and $X$ characterizes the experiment: $X=0.25$ for BBO and $X=100$ for DECIGO.  
In Figure 2c we present the best-case reconstruction including a direct detection of $n_T$ at DECIGO-precision with a fiducial
value of $n_T = -r/8 = -0.01875$, in agreement with all three consistency conditions.  
We find that the curvaton models (black) are almost as well constrained as canonical single field inflation (magenta), while for
DBI (blue) we find
$V''_{\rm DBI} \approx V''_{\rm csf}$ and $V'_{\rm DBI} \approx 2V'_{\rm csf}$.   
In addition, with a measurement of $n_T$ the zoology can be partially recovered 
compared to the
worst-case degeneracy, Figure 2b.  In the event of a future detection with DECIGO together with a
moderate amplitude of tensors, $r = 0.15$, the zoology possesses regions occupied uniquely by small
field, large field, and hybrid models, shown in Figure 2d.  Constraints on models with smaller fiducial $r$ are also improved although to a
lesser degree.  We note that measurements of $n_T$ at BBO-precision also offer
improvements over the worst-case degeneracy, but for brevity we present only the best-case reconstruction here (see
\cite{debp10a,debp10b}).

In conclusion, while we may never know the true underlying theory of inflation, we have found that it is possible to                                                  
vastly improve our understanding of the inflaton potential.
%
%
%
%

\acknowledgments
The work of DAE is supported in part by the World Premier International Research 
Center Initiative (WPI Initiative), MEXT, Japan and by a Grant-in-Aid for Scientific Research 
(21740167) from the Japan Society for Promotion of Science (JSPS), and by the Arizona State University 
Cosmology Initiative.


\begin{thebibliography}{99}
\bibitem{note2}
In addition to Planck, we consider constraints expected from CMBPol in the more extensive analyses of \cite{debp10a,debp10b}.
\end{thebibliography}
\end{document}